\documentclass[twocolumn]{cinc}
\usepackage{graphicx}
\usepackage{tikz}
\usetikzlibrary{shapes,arrows,positioning,calc}
\usepackage{pgfplots}
\pgfplotsset{compat = 1.15}
\usepgfplotslibrary{groupplots}
\usepackage{amsmath}
\usepackage{mathabx}
\usepackage{xcolor}
\usepackage[absolute,overlay]{textpos}
\begin{document}
\bibliographystyle{cinc}

\begin{textblock*}{18cm}(2cm,30cm)
This work has been submitted to the IEEE for possible publication. Copyright may be transferred without notice, after which this version may no longer be accessible.
\end{textblock*}

\title{Combining Scatter Transform and Deep Neural Networks for Multilabel Electrocardiogram Signal Classification}

\author { Maximilian P~Oppelt$^{1}$, Maximilian Riehl$^{1}$, Felix P~Kemeth$^{2}$, Jan Steffan$^{1}$
\ \\ 
$^1$Department of Image Processing and Medical Engineering, Fraunhofer IIS \\
$^2$Department of Chemical and Biomolecular Engineering, Whiting School of Engineering, Johns Hopkins University}

\maketitle

\begin{abstract}
  An essential part for the accurate classification of electrocardiogram (ECG) signals
  is the extraction of informative yet general features, which are able to discriminate diseases. Cardiovascular abnormalities manifest themselves in features on different time scales:
  small scale morphological features, such as missing P-waves,
  as well as rhythmical features apparent on heart rate scales.

  For this reason we incorporate a variant of the complex wavelet transform,
  called a scatter transform, in a deep residual neural network (ResNet).
  
  The former has the advantage of being derived from theory, making it well behaved under certain transformations of the input.
  The latter has proven useful in ECG classification, allowing feature extraction and classification to be learned in an end-to-end manner.
  
  Through the incorporation of trainable layers in between scatter transforms, the model gains the ability to combine information from different channels,
  yielding more informative features for the classification task and adapting them to the specific domain. 

  For evaluation, we submitted our model in the official phase in the
  PhysioNet/Computing in Cardiology Challenge 2020.
  Our (Team Triage) approach achieved a challenge validation score of 0.640, and full test score of 0.485, placing us 4th out of 41 in the official ranking.
\end{abstract}

\section{Introduction}
Electrocardiography is a non-invasive technique to record the electrical activity of the
heart using a set of electrodes.
It captures small electrical changes on the skin caused by cardiac
depolarization and repolarization during each heart cycle.
Pathological changes, including arrhythmias,
may alter the electrical properties of a heart beat and thus cause changes
in the recorded electrocardiogram.

For the task of automatically classifying these ECG signals,
older methods rely on handcrafted features constructed by domain experts,
as well as on information extracted via classical signal processing.
More recently machine learning has been demonstrated to be a competitive alternative.


Both fields have their respective advantages and weaknesses. Our approach aims to improve upon the state of the art by combining a residual network (ResNet) with a signal processing method called scatter transform. We view this as constructing an intermediate design which can be interpreted as a well understood classical method augmented with the ability to learn.

By introducing more inductive bias into the deep net we hope to reduce problems associated with overfitting. 


The dataset is provided by the Physionet/CinC2020 challenge consisting out of 12 lead labeled ECG recordings, described in \cite{physionet2020}.

\section{Methodology}


We introduce a deep neural network which uses a modified ResNet as the encoder for the data. The modification replaces some layers in the bottleneck blocks of the ResNet with scatter transforms. The encoder module is followed by a multi-head self attention layer, before feeding its output into a stack of fully connected layers for classification
(see Table~\ref{tab:network} for a summary of the architecture).
To optimize the network towards the challenge metric during training,
we employed a differentiable version of the metric as the loss function.

\subsection{Preprocessing}
The Physionet/CinC2020 dataset contains 43101 annotated recordings of
different lengths, labeled with one or more of distinct 111 classes.
The challenge evaluation metric contains a subset of only 27 classes.
We dropped records that only consist of classes not intersecting with the evaluation metric.
In addition we joint classes with identical scores, yielding a 24 class problem. Recordings sampled with a frequency different from $500\, \mathrm{Hz}$ were resampled in order to match that sampling rate.

We chose 10240 samples ($20.48\,\mathrm{s}$) for the size of the prepossessing window.
ECGs longer than the given input window where split into equally sized recordings.
The input signals were normalized to have zero mean and unit variance and then passed through the arctan function.
This reduces the size of the R-peaks relative to the rest of the morphology in order to prevent them from dominating the feature extraction.

Using the afore mentioned approach of input data preprocessing we get 44582 data points for training, 4458 for validation and 499 holdout records.
After only keeping the records used for the final submission with the 24 evaluated classes, our train/validation/holdout split consists of 37281/3720/412 records.

\subsection{Augmentation}
To reduce overfitting we applied specialized data augmentation techniques.
First we randomly add power noise with frequencies around $50\,\mathrm{Hz}$ and secondly, Gaussian noise with zero mean and a standard deviation of $0.08$ is added.
We also introduce a sinusoidal drift with random phase, frequency and amplitude,
that resembles a baseline drift.

The selected input window of our network is 5120 samples ($10.24\,\mathrm{s}$) long.
For each data point we randomly sample the window location from the pre-processed signal.
Recordings shorter than the given input window are padded with zeros.

\subsection{Scatter Transform}
\label{sec:scattering}
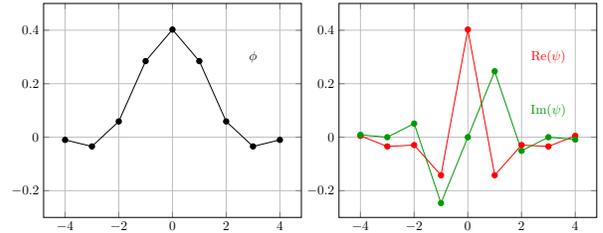
\begin{figure}[hptb]
    \begin{tikzpicture}[scale = 0.5]
    
        \begin{groupplot}[
        group style = {group size = 2 by 1,},
        xmajorgrids=true,
        ymajorgrids=true,
        ymin = -0.3,
        ymax = 0.5,
        ]
            \nextgroupplot
            
        	\addplot [color=black, mark = *] coordinates {
            	( -4 , -0.010110028629970187 )
                ( -3, -0.03451779686442458 )
                ( -2 , 0.05892556509887895 )
                ( -1 , 0.2845177968644246 )
                ( 0 , 0.40236892706218247 )
                ( 1 , 0.2845177968644246 )
                ( 2 , 0.05892556509887895 )
                ( 3 , -0.03451779686442458 )
                ( 4 , -0.010110028629970187 )
        	};
        	
        	\node[] at (axis cs: 3.,0.3){$\phi$};
    
        	\nextgroupplot
        	
        	\addplot [color=red, mark = *] coordinates {
            	( -4 , 0.005055014314985086 )
                ( -3 , -0.03451779686442458 )
                ( -2 , -0.029462782549439504 )
                ( -1 , -0.14225889843221223 )
                ( 0 , 0.40236892706218247 )
                ( 1 , -0.14225889843221223 )
                ( 2 , -0.029462782549439504 )
                ( 3 , -0.03451779686442458 )
                ( 4 , 0.005055014314985086 )
        	};
        	\addplot [color=green!60!black, mark = *,] coordinates {
                ( -4 , 0.008755541626542171 )
                ( -3 , -8.45442188872722e-18 )
                ( -2 , 0.05103103630798286 )
                ( -1 , -0.24639963991337221 )
                ( 0 , 0.0 )
                ( 1 , 0.24639963991337221 )
                ( 2 , -0.05103103630798286 )
                ( 3 , 8.45442188872722e-18 )
                ( 4 , -0.008755541626542171 )
        	};
        	
        	\node[color=red] at (axis cs: 3.,0.3){$\mathrm{Re}(\psi)$};
        	\node[color=green!60!black] at (axis cs: 3.,0.1){$\mathrm{Im}(\psi)$};

        	\end{groupplot}
    \end{tikzpicture}
    \caption{Low-pass filter ($\phi$, left) and high-pass filter ($\psi$, right) with the numerical values given in Table~\ref{tab:wav_coeff}.}
    \label{fil_img}
\end{figure}

The task of robust time series classification crucially depends on the underlying features that are fed into the classifier.
The usual approach in machine learning is to make an educated guess about the appropriate architecture and learn the weights of the classifier and of the feature extractors end-to-end.\\
The so called scatter transform~\cite{invariantScattering} offers a principled alternative to this process by providing us with features from a fixed convolutional network, the structure of which is derived from theory without any trainable parameters.\\
The local structure inherent to time series data is the reason for the
widespread use of convolutional architectures in various types of classification tasks.
Their usefulness stems from their ability to deal with variability of the signal due to translations in time.\\
The scatter transform extends this by putting a \mbox{Lipschitz} constraint on the network in order to make the features change smoothly with local deformations of the input. This results in a features space in which the euclidean distance is able to capture, as we argue, a more useful concept of similarity between two input signals. As it turns out these conditions lead to a particular choice for the filters and the non-linearity~\cite{bruna:pastel-00905109}.\\
\\
A single layer of the scatter transform is similar to the discrete wavelet transform.
Each channel of the input is convolved separately by a high-pass wavelet $\psi$ and a low-pass wavelet $\phi$. Both filters are computed with a stride of 2, resulting in an output which is sub-sampled in time, but has double the number of channels.
In particular, each channel $x$ of the signal is mapped to
\begin{align}
    \label{equ:scatteroperation}
    x\mapsto \mathrm{stack}\begin{pmatrix}
    \left(x \ast \phi \right) \downdownarrows 2 \\
    \left|x \ast \psi \right| \downdownarrows 2 \end{pmatrix}
\end{align}
where $\ast$ indicates convolution in time, $\left| \cdot \right|$ the absolute value and $\downdownarrows 2$ downsampling by a factor of 2.
It is necessary that the high-pass filter is approximately analytic, in order for it to have a smooth magnitude. A possible choice for the filters is displayed in Figure \ref{fil_img}, with the coefficients being summarized in Table~\ref{tab:wav_coeff}.

\begin{table}[hbtp]
    \centering
    \begin{tabular}{c | c c}
          $\phi$ & $\psi$ &\\
          \hline
        -0.0101100286 &  0.0050550143 &+ 0.0087555416j \\
        -0.0345177968 & -0.0345177968 &+ 0j \\
         0.0589255650 & -0.0294627825 &+ 0.0510310363j \\
         0.2845177968 & -0.1422588984 &- 0.2463996399j \\
         0.4023689270 &  0.4023689270 &+ 0j \\
         0.2845177968 & -0.1422588984 &+ 0.2463996399j \\
         0.0589255650 & -0.0294627825 &- 0.0510310363j \\
        -0.0345177968 & -0.0345177968 &+ 0j \\
        -0.0101100286 &  0.0050550143 &- 0.0087555416j \\
    \end{tabular}
    \caption{Coefficients of the discrete low-pass ($\phi$) and complex high-pass ($\psi$) wavelets used in the scatter transform.}
    \label{tab:wav_coeff}
\end{table}

The low-pass filtered signal essentially gives a low resolution representation of the input.
Following the Nyquist theorem we are allowed to reduce the number of sample points to represent the signal. This path achieves stability with regard to small local deformations by averaging out and thus removing detailed information.
\\
The main difference compared to a regular wavelet transform consists in the application of the modulus function after the high-pass wavelet.
It removes the phase, which encodes small local translations, and makes the output real valued.
This operation computes the envelope of the filter response which varies more slowly with local translations than the rapidly oscillating phase.
The purpose of this computation is to capture the information about the presence of high frequency oscillations that is lost in the low-pass.\\
\\
It is of note that the output of a scatter layer has the same number of variables as the input. With the additional condition that the input is purely real,
it turns out that the scatter transform, despite loosing the phase, is invertible and thus conserves information~\cite{DeepScatteringSpectrum}.\\

\subsection{Model Architecture}
Our baseline network consists of a ResNet encoder followed by a self attention block \cite{vaswani2017attention} and a fully connected classifier. To stabilize training and increase accuracy the swish activation function $x \mapsto x \cdot \mathrm{sigmoid(x)}$ is employed for the ResNet blocks and the intermediate layers of the classifier ~\cite{ramachandran2017searching}.
To reduce overfitting we use Dropout with dropout probabilities of 0.25 in between the layers of the classifier and batch normalization after each convolution. Finally a sigmoid function is used to transform the logits from the output into probabilities for the given mulitclass/mulitlabel task.
Thresholding of the probabilities leads to binary class predictions.

\begin{table}[hbtp]
    \begin{tabular}{l|l|l|l}
    layer name  & input & output & parameter                                                                    \\ \hline
    conv1d.1    & 12    & 24     & kernel: 7, $\downdownarrows 2$                                                         \\ \hline
    maxpool.1   & 24    & 24     & kernel: 3, $\downdownarrows 2$                                                         \\ \hline
    residual.1.x  & 24    & 48     & $\begin{bmatrix} \mathrm{conv}1, 3 \\ \mathrm{conv}3, 6 \\ \mathrm{conv}1, 48  \end{bmatrix}\times3$    \\ \hline
    residual.2.x  & 48    & 96    & $\begin{bmatrix} \mathrm{conv}1, 6 \\ \mathrm{conv}3, 12\\ \mathrm{conv}1, 96  \end{bmatrix}\times4$, $\downdownarrows 2$  \\ \hline
    residual.3.x  & 96   & 192    & $\begin{bmatrix} \mathrm{conv}1, 12 \\ \mathrm{conv}3, 24 \\ \mathrm{conv}1, 192  \end{bmatrix}\times6$, $\downdownarrows 2$ \\ \hline
    residual.4.x  & 192   & 384    & $\begin{bmatrix} \mathrm{conv}1, 24 \\ \mathrm{conv}3, 48 \\ \mathrm{conv}1, 384  \end{bmatrix}\times3$, $\downdownarrows 2$ \\ \hline
    conv1d.2    & 384   & 96     & kernel: 1, $\downdownarrows 2$                                                         \\ \hline
    attention.1 & 96    & 96     & attention heads: 12                                                          \\ \hline
    avgpool     & 96    & 96     & adaptive output size: 8                                                      \\ \hline
    fc.1 & 770   & 256    &                                                                              \\ \hline
    fc.2 & 256   & 24     &                                                                              \\ 
    \end{tabular}
    \caption{Summary of our model architecture}
    \label{tab:network}
\end{table}

\subsubsection{ResNet}
The core concept behind the design of residual networks is to express its computation as a series of perturbations to the identity function \cite{he2016deep}. One such block of the ResNet can be expressed through a subnetwork $\mathcal{F}$:
$\pmb{y} = \pmb{x} + \mathcal{F} \mathnormal{\left(\pmb{x}\right)}$
With the goal of saving computational resources, $\mathcal{F}$ can be made to be in the shape of a bottleneck, meaning for its internal processing it first projects the input to a lower dimensional space. At the output it projects back to match the shape with the input again.\\
In our case the bottleneck consists of three 1D convolutions with kernel sizes 1/3/1. By also applying a projection $W_d$ within the skip connection, typically implemented as a convolution of kernel size 1 and stride 2, it is possible for the residual block to change the temporal as well as the channel dimension of its output:
$\pmb{y} = W_d \pmb{x} + \mathcal{F} \mathnormal{\left(\pmb{x}\right)}$




\subsubsection{Scatter blocks}
Despite the afore mentioned properties of the scatter transform, it is not able to capture all relevant variability in the ECG-classes. Some properties of the data are domain specific and need to be learned. Among these are for example interaction between channels, which the scatter transform does not address since it processes all of them separately.
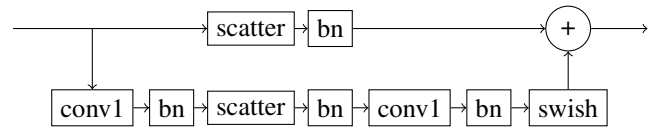
\begin{figure}[htpb]
    \tikzstyle{block} = [draw, fill=white, rectangle, 
        minimum height=0.6em, minimum width=0.5em , align = center]
    \tikzstyle{sum} = [draw, fill=white, circle]
    \tikzstyle{input} = [coordinate]
    \tikzstyle{output} = [coordinate]
    \tikzstyle{pinstyle} = [pin edge={to-,thin,black}]
    
    \begin{tikzpicture}[node distance=3em,]
    
        \node [input](input) {};
        \node [input, left of = input](source){};

        \node [block, below of =  input](s2){conv1};
        \node [block, right of = s2](bn2){bn};
        
        \node [block, right of = bn2](s3){scatter};
        \node [block, right of = s3](bn3){bn};
        
        \node [block, right of = bn3](s4){conv1};
        \node [block, right of = s4](bn4){bn};
        
        \node [block, right of = bn4](a1){swish};
        
        \node [sum, above of = a1](sum){+};

        \node [block, above of = s3](s1){scatter};
        \node [block, right of = s1](bn1){bn};

        \node [output, name = output, right of = sum] {};

        \draw[->](input) -- (s1);
        \draw[->](s1) -- (bn1);
        \draw[->](bn1) -- (sum);

        \draw[->](input) -- (s2);
        \draw[->](s2) -- (bn2);
        
        \draw[->](bn2) -- (s3);
        \draw[->](s3) -- (bn3);
        
        \draw[->](bn3) -- (s4);
        \draw[->](s4) -- (bn4);
        
        \draw[->](bn4) -- (a1);
        \draw[->](a1) -- (sum);

        \draw[-](source) -- (input);
        \draw[->](sum) -- (output);

    \end{tikzpicture}

    \caption{The scatter blocks use scatter layers in place of convolutions with stride 2. In the reference bottleneck blocks there is a convolution with kernel size 1 and stride 2 in the skip connection when downsampling is performed.}
    \label{fig:scatter resnet}
\end{figure}

For this reason we combine a ResNet architecture with the scatter transform. One obvious way to do this is to use the scatter layer for temporal downsampling, in particular for convolutional layers with stride equal to 2.

The scatter layer (equation \ref{equ:scatteroperation}) acts as a drop in replacement for the projection $W_d$ of the skip connections as well as the convolutional layers with temporal downsampling in the residual blocks $\mathcal{F}$, see Figure \ref{fig:scatter resnet}.

The controlled downsampling suppresses aliasing, which as we argue leads to a better representation of the signal inside the network. The reference ResNet implementation has stride 2 only in layers residual.2.1, residual.3.1 and residual.4.1 (cf. Table \ref{tab:network}). We replace these with scatter blocks (cf. Figure \ref{fig:scatter resnet}).

\subsubsection{Attention}
In ECG classification it is common to capture the sequential information using recurrent layers \cite{hong2020opportunities}. In our approach we opt to use an attention mechanism to capture the temporal dependencies instead. We implemented a multi head attention block following \cite{vaswani2017attention}. Our multi-head multiplicative attention block has 32 input channels and attends using 4 heads. We employ positional encoding as described in \cite{vaswani2017attention}. The motivation behind this is that we want to disproportionately weigh parts that are indicative for a particular disease. The attention layer is able to focus on these important regions.

\subsubsection{Cost-function}
The metric provided by the challenge organisers assumes discrete class labels for evaluation, which makes it unsuitable for direct optimization by gradient descent. We use a workaround by constructing a differentiable analog for the logical OR in the normalization constant $n$, the purpose of which is to discourage simply classifying every label as true in all instances.\\
The challenge loss incorporates a matrix $W$ to account for how undesirable choosing the wrong classification $p$ for a given ground truth $t$ is. The most common cost function for independent boolean classes is the binary cross entropy. In fact we found that a 1 to 1 weighting of both losses performed better than each one by itself:
\begin{align}
    n &= \sum_{i=0} (t_i + p_i - t_i \cdot p_i) \approx \sum_{i=0} (t_i \lor p_i)\\
    L &= - t ^ T \cdot \log(p) - (1-t) ^ T \cdot \log(1-p)  - \tfrac{t^T \cdot W \cdot p}{n}
\end{align}
with the first two terms on the right hand side in equation 3 being the binary cross entropy loss and the last term the differentiable analog to the challenge metric.

\subsubsection{Training}
Our implementation uses the Adam optimizer with a learning rate of $0.003$ and learning rate reduction whenever the training error does not decrease for 12 epochs. We train both models for a maximum of 256 epochs and select the model that performs the best on the validation set. The batch size for training and validation is 256.

\section{Discussion}
By comparing the standard ResNet with a version augmented by scatter layers the metric on our holdout dataset increased from 0.682 to 0.724.
We (Team Triage) expect this increase in performance to be due to faster convergence, stemming from the reduced number of parameters and the Lipschitz properties of the scatter layers.

Despite using fewer parameters in our scatter ResNet (166504 parameters) it outperforms the default ResNet bottleneck network (214957 parameters) for the described setup. This indicates that the modifications provide inductive bias that is suitable for the given task.

Future research needs to analyze the properties of the scatter layer augmented network in greater detail, in particular with regards to the problem of overfitting.




\balance

%


%

\section*{Acknowledgments}  
%
This work was supported by the Bavarian Ministry for Economic Affairs, Infrastructure, Transport and Technology through the Center for Analytics-Data-Applications (ADA-Center) within the framework of “BAYERN DIGITAL II”.
\\
This work was supported by Matthias Struck, Deputy Head of Department Image Processing and Medical Engineering, by providing a deep learning cluster and financial resources.

\bibliography{ms}


\begin{correspondence}
Matthias Struck, Deputy Head of Department Image Processing and Medical Engineering
Fraunhofer IIS,
Am Wolfsmantel 33,
91058 Erlangen, Germany,
matthias.struck@iis.fraunhofer.de

\end{correspondence}

\end{document}